\documentstyle[11pt]{article}
\input epsf

\newcommand{\nl}{\newline}
\newcommand{\al}{{\alpha}}

\newcommand{\pa}{{\partial}}
\newcommand{\beq}{\begin{equation}}
\newcommand{\eeq}{\end{equation}}
\newcommand{\beqa}{\begin{eqnarray}}
\newcommand{\eeqa}{\end{eqnarray}}
\newcommand{\ben}{\begin{enumerate}}
\newcommand{\een}{\end{enumerate}}
\newcommand{\bi}{\begin{itemize}}
\newcommand{\ei}{\end{itemize}}

\newcommand{\lsim}{\hbox{ {\raisebox{0.06cm}{$<$} \raisebox{-0.14cm}{$\!\!\!\!\!\!\!\!\: \sim$}} } }
\newcommand{\rsim}{\hbox{ {\raisebox{0.06cm}{$>$} \raisebox{-0.14cm}{$\!\!\!\!\!\!\!\!\: \sim$}} } }

 \title{Statics and Fast Dynamics of Nanomagnets with Vortex Structure}

\author{ R.\ H\"ollinger{\thanks{Corresponding author,
 e-mail rainer.hoellinger@physik.uni-regensburg.de}, A.\ Killinger, U.\ Krey}
\\  Inst.\ f\"ur Physik II, Universit\"at Regensburg, 93040 Regensburg,
Germany
  }

\date{v1: April 23, 2002; v3: Nov.\ 11, 2002, accepted by JMMM}
\input epsf
\begin{document}

\large
\maketitle
\begin{abstract}

\noindent Within the framework of the Landau-Lifshitz-Gilbert equation,
using permalloy parameters, we study the statics and dynamics of flat
circular magnetic nano-structures with an in-plane magnetic vortex
configuration, putting particular emphasis on the {\it vorticity} of the
magnetic state and on the (perpendicular) {\it polarisation} of the vortex
center, which may be shifted with respect to the center of the circle. These
binary degrees of freedom can in principle be used to manipulate {\it two
independent bits} of information.

Studying switching processes induced by in-plane and out-of plane field
pulses we find that it is possible to switch the vorticity of the
magnetic dot on a time scale of $40$ ps in strong enough and short enough
 perpendicular external field pulses ($B_z^{{\rm ext}}\approx 0.5$ T,
duration $\approx 40$ ps). But for realistically small values of the Gilbert
damping, only the {\it vorticity} can be switched this fast, and it turns
out that it is better to dismiss the center of the circle totally,
concentrating on flat 'nano-rings' with an inner radius $R_1$ and an outer
radius $R_2$.  On these 'nano-rings' the vortex state is more stable, and
with respect to the switching of the vorticity these structures have similar
properties as circular dots.

\end{abstract}
\vspace*{0.5cm}
PACS numbers/keywords:

\noindent
75.75.+a Magnetic properties of nanostructures;
75.40.MG Numerical simulation studies;
75.40 Gb Dynamic properties\nl
85.70-d  Magnetoelectronics

\section{Introduction}

 Recently there is a strong interest in using small magnetic structures as
storage elements and for fast magnetoelectronic computation (MRAMs,
\cite{GPrinz}). R.\ Cowburn, \cite{RCowburn}, has studied experimentally and
by simulation arrays of flat magnetic structures with circular or square
individual geometries, and M.\ Schneider {\it et al.},
 \cite{Schneider2001}, have considered more general geometries of the dots.
The present state of the art concerning the static behaviour of magnetic
nano-dots and their arrays is contained in the very recent papers of Ross
{\it et al.}, \cite{Ross}, Metlov and Guslienko, \cite{Metlov}, and
Guslienko {\it et al.}, \cite{Guslienko2002,Usov}. Concerning the dynamics we
explicitly mention the even more recent papers of Gerrits {\it et al.},
\cite{Gerrits}, and Novosad {\it et al.}, \cite{Novosad}, and the references
therein.

If for a given thickness  the radius $R$ of the flat dot is smaller than a
critical value $R_c$ (see below), then the dot is in an in-plane
single-domain state, whereas for $R > R_c$, the dot is magnetically in a
{\it vortex state}, see
\cite{RCowburn}.

Below, for flat dots, i.e.\ for small values of their {\it aspect ratio}
$a:=({t_h}/{R})\ll 1$, where $t_h$ denotes the {\it thickness} of the dot,
at first (for given $R$) the critical thickness $t_c$ is calculated for
structures with permalloy material parameters, and a considerable stability
range for the magnetic vortex state is obtained even in rather tiny
nano-structures. By application of a sufficiently small in-plane static
magnetic field the magnetic state of the dot does not change topologically,
i.e.\ the vortex center shifts, however the vorticity is still well-defined,
and also the central direction of the magnetization (so-called ''central
polarisation'' $\pm$) is unchanged. At the vortex center and in the region
nearby, the magnetization (which must be constant in magnitude) is
out-of-plane, whereas far from the center it is of course in-plane.

Therefore the magnetic vortex state of the dot has a fourfold topological
degeneracy, since in principle one can measure (and {\it flip})
independently both the {\it vorticity} ($r/l$) of the state and the {\it
central polarisation} ($+/-$) of the vortex; {\it i.e.\ one has two bits per
dot for a 'vortex-state dot', instead of only one bit for 'single-domain
dots'}.

At present, the {\it vorticity} of the dot can be measured e.g.\ by electron
microscopy in the Lorentz mode while the {\it central polarisation} can be
measured e.g.\ by magnetic force microscopy, \cite{JRaabe}: Of course, for
the applications one should find more convenient methods; but this task is
not considered here. In any case, the possibility of storing and switching
two independent bits of information instead of only one, and the fact that
one does not rely on the extremely small single-domain dots, gives a strong
additional motivation to study the statics and dynamics of circular dots in
the vortex state very thoroughly.

In the following chapter 2 we introduce the basic micromagnetic tools for
such a study, the Landau-Lifshitz free energy functional and the dynamical
Landau-Lifshitz-Gilbert equation, \cite{Gilbert}. In chapter 3, the
behaviour of flat circular dots under static external magnetic {\it
in-plane} fields is considered, with particular emphasis on the transitions
between the vortex state and the in-plane single-domain state, whereas in
the next chapter 4 we consider the dynamics, both for in-plane fields and
perpendicular field-pulses. The switching-dynamics is exemplified by some
kind of ''movie representation'' for typical cases, which show that the time
scale, $\tau\sim 40$ ps, for the precessional switching of the vorticity by
strong perpendicular field pulses, $H_z^{\rm ext}\sim 5000$ Oe over this
duration, is extremely short, whereas the polarisation of the vortex core
cannot be reliably switched at all on these time scales, see below.
Therefore at the end, it is natural to skip the vortex center totally by
considering in chapter 5 flat {\it nano-rings}, i.e.\ circular structures
with an inner radius $R_1$ and an outer radius $R_2$, which are both small,
i.e.\ of the order of 50 to 150 nm, but large compared to the thickness
$t_h$. We find that in this case the stability of the information is
significantly improved and the switching dynamics is as fast as before.

\section{Basic equations}

Let us write the magnetic polarisation $\vec J(\vec r,t)$ as $\vec J(\vec
r,t)
=J_s\cdot\vec\al (\vec r,t)$, where $J_s = |\vec J|$ is the constant {\it magnitude}
 and $\vec
\al (\vec r, t)$ the position- and time-dependent {\it direction}
 of the magnetic polarisation. We neglect in the following the small
(in-plane) uniaxial magnetic anisotropy of Py, taking only
 the exchange interaction, the magnetic dipole-dipole interaction, and the 
Zeeman interaction (i.e.\ the interaction with the external magnetic field)
 into account. As a consequence, the Free Energy of the magnetic system $K$
considered is obtained by minimization of the following ''Landau-Lifshitz
functional'', see e.g.\ \cite{Aharoni,Hubert}:
\beq\label{eqLandauLifshitz}
F_{LL}:=\int\limits_{V(K)}\,{\rm d}V\cdot \left\{A\,\sum\limits_{i=1}^3
 (\frac{\pa \vec\al}{\pa x_i })^2 -J_s\cdot (\frac{1}{2}\,\vec H^{\rm Dip}\cdot\vec\al
+\,\vec H^{\rm ext}
\cdot\vec\al  ) \right\}
\eeq

Here $V(K)$ is the integration volume of the magnetic system $K$, $A$ is the
''exchange constant'', $J_s$ is the above-mentioned saturation polarisation
(we use $A\cong 1.3\times 10^{-6}$ erg/cm and $J_s\cong 1$ Tesla for Py,
unless otherwise stated); $\vec H^{\rm ext}$ is the external magnetic field
and $\vec H^{\rm Dip}$ the internal
 dipolar field
 (also called: ''stray-field'') which is generated by the magnetic moments
of the system according to the magnetostatic equation
$
\vec H^{\rm Dip}(\vec r,t)=-{\rm grad}\,\phi_m(\vec r,t),\quad{\rm with}
$
$\nabla^2\phi_m=\frac{1}{\mu_0}\,{\rm div}\left (\,J_s\cdot\vec\al
\right )
$
and the corresponding  boundary and interface constraints.  (Note that at
this place, even on a ps time scale, we can safely use the magnetostatic
equations without taking into account retardation effects, since e.g.\ in
the time of $\tau =1$ ps a vacuum electromagnetic wave would traverse a
distance of 300 $\mu\,m$, which is far beyond the length scales of the
structures considered, \cite{REM1}.)

The dynamic equation of motion corresponding to (\ref{eqLandauLifshitz})
is
\beq
\frac{{\rm d}\vec\al}{{\rm d}t}=-\gamma_0\,\vec\al \times\vec H^{\rm eff}\,,
\eeq
\vglue 0.2 truecm

 \noindent where $\gamma_0 =g\frac{|\mu_B|}{\hbar}$ is the  ''{\it
gyromagnetic ratio}'' ($\mu_B$ is Bohr's ''magneton'', $\hbar$ Planck's
constant divided by $2\pi$, and $g$ the  Land\'e factor,
$g = 2$ for spin magnetism, i.e.\ in this case $\gamma_0$ has the value
$1.176\cdot 10^7$ s$^{-1}$/Oe), whereas $\vec H^{\rm eff}$ is the 
space- and time-dependent effective magnetic field,
around which the magnetic polarisation precesses, namely \beq
\label{eqEffectiveField} \vec H^{\rm eff} =\vec H^{\rm ext}+\vec H^{\rm
Dip}+\frac{2A}{J_s}\,\nabla^2\vec\al -
\frac{\al_G}{\gamma_0}\cdot\frac{{\rm d}\vec\al}{{\rm d}t}\,. \eeq

Here -- as already mentioned -- $\al_G$ is Gilbert's  damping constant,
\cite{Gilbert}; i.e.\ the last term in equation (\ref{eqEffectiveField})
describes an isotropic velocity dependent phenomenological damping leading
 to a gradual decay of the precession amplitude. In principle, and
particularly near interfaces, this damping should be anisotropic, see e.g.\
\cite{LosAlamos1,Fraermann}, but in the following this damping anisotropy is
neglected, and also the thermal fluctuation fields, which would be
proportional to
$(T\cdot\al_G)^{\frac{1}{2}}$, are neglected as usual (but see the recent
papers of Brown {\it et al.} and Nowak {\it et al.}, \cite{Brown,Nowak}).
Also deviations from cubic symmetry of the magnetic material, which would
lead to a more general density of the exchange energy, namely
$\sum\limits_{i,j=1}^3 A_{ij}\,
 \frac{\pa\vec\al}{\pa x_i}\cdot\frac{\pa\vec\al}{\pa x_j}$, are neglected.

\section{Static behaviour of a magnetic dot: Vortex state versus
in-plane single-domain state}
\subsection {Vanishing external field}
At first we  consider the static behaviour of a flat circular magnetic dot
for vanishing external field; in particular we consider the phase boundary
between the vortex state and an in-plane single-domain state,
\cite{REMoutofplane}, minimizing the
''Landau-Lifshitz Functional'' (\ref{eqLandauLifshitz}) with the {\it
ansatz}
\beqa  \label{eq4} \al_x(r) &=& -\sin\varphi (r)\cdot( 1 -\al_z^2(r))^{\frac{1}{2}}\cr
\al_y(r) &=& +\cos\varphi (r)\cdot( 1 -\al_z^2(r))^{\frac{1}{2}} \eeqa
\noindent Our special ansatz for the 'profile function' $\al_z(r)$ is an
improved version of an approximation by Feldtkeller and Thomas,
\cite{Feldtkeller}), led by the following considerations:

For $r\ll l_m$, where  $l_m :=(\frac{2A\mu_0}{J_s^2})^{\frac{1}{2}}$ is the
'magnetic exchange length', one expects the behaviour $\al_z(r)=1-{\rm
const.}\cdot{r^2} -...$ (where the dots denote higher-order terms); beyond
that range one will observe an exponential spatial dependence with decay length
$l_m$, while asymptotically for $r\gg l_m$ some kind of 'Mexican Hat'
behaviour will apply, i.e.\ after the pronounced exponential decay
$\al_z(r)$ should become negative and approach the $r$-axis from negative
values.  However, this negative part should be proportional to the magnetic
moment of the central part of the dot, i.e.\ $\sim J_s\pi l_m^2 t_h$,
and therefore it vanishes $\propto t_h$ for very flat dots.

  (Our {\it ansatz} for $\al_z(\vec r)$, see below, is more accurate than that of
eq.\ (5) of Usov and Peschany, \cite{Usov}, who neglect the magnetostatic
energy generated by the vortex core completely to obtain their simplified
profile function, which contains a singularity of the derivative at the
so-called core radius and is constant for larger $r$, i.e.\ it does not
contain the tail nor the negative part of the profile. Our calculation is
also more accurate, see the results, than that of Guslienko and Metlov,
\cite{Guslienko2002}, and Guslienko {\it et al.}, \cite{Guslienkoetal}, who
take the magnetostatic field produced by the vortex core into account, but
use basically the simplified core profile of \cite{Usov} instead of
the more correct one of (\ref{ansatz2}) below.)

In any case, as a systematic approximation, expanding the ansatz of
Feldtkeller and Thomas, \cite{Feldtkeller}, we use in eq.\ (\ref{eq4}):
\beq\label{ansatz2}\al_z(r)\equiv c\cdot e^{-\frac{r^2}{l_m^2}}+(1-c)\cdot
e^{-\frac{r^2}{4l_m^2}},\eeq
where $c$ depends on the thickness of the sample and is
obtained by an energy-fitting procedure leading to
\beq \label{eqC} c\approx -0.188 + 0.708\cdot (1+0.2\cdot T_h+0.0278\cdot
T_h^2)^{-1}\,.\eeq
Here $T_h :=\frac{t_h}{l_m}$ is the reduced thickness of our structures.
 Note that $c\approx 0.5$ for very flat dots ($t_h\ll R$). (Eqs.\
(\ref{ansatz2}) and (\ref{eqC}) are of course only approximately valid,
since 'Mexican hat profiles', which one would get for thicker dots, are
explicitly excluded for $-0.188 \le c \le 0.52$.)

In fact, the gross behaviour of the profile function for our flat structure
is given by the above-mentioned ansatz (\ref{ansatz2}):
 This is exemplified in Fig.\ 1, where the results of a numerical simulation
of $\al_z(r)$, performed with the OOMMF program, \cite{OOMMF}, for flat
circular permalloy structures with $R=175.5$ nm and a thickness of
$t_h=6$ nm, are compared with (\ref{ansatz2}) and (\ref{eqC}). Details
are given in Fig.~1a and in the corresponding figure caption. In Fig.~1b an
effective radius $R_{\rm eff}$ of the central region of the vortex is
presented as a function of the reduced thickness $T_h$. This effective
radius $R_{\rm eff}$ is defined by the equation $R_{\rm
eff}=\frac{\pi}{2}\cdot\frac{1}{\left (\frac{{\rm d}\theta}{{\rm
d}r}|_{r=0}\right )}$, taking into account the fact that the polar angle
$\theta$ of the magnetic polarisation is zero rsp.\ $\frac{\pi}{2}$ in the
vortex center rsp.\ far from it. With (\ref{ansatz2}) one gets $R_{\rm
eff}\cong\frac{\pi}{\sqrt{2+6\cdot c}}\cdot l_m$, which is plotted in
Fig.~1b.  This means that $R_{\rm eff}$ is approximately between $1.5\,l_m$
and $3\,l_m$ depending on the thickness $t_h$.

Our ansatz $\al_z(r)$, eq.\ (\ref{ansatz2}), for the above-mentioned profile
function leads to {\it analytical} formulae for the exchange energy density
$ f^{\rm exch}=\frac{A}{V} \int\limits_V\,\sum\limits_{i=1}^3\,
(\frac{\pa\vec \al}{\pa x_i })^2\,{\rm d}V $, and we obtain for thin dots
with $c=0.5$ approximately: $f^{\rm exch}\approx \frac{2A}{R^2}\cdot
[2.67 + \ln \frac{R}{5l_m}]$.

 For  circular single-domain dots with homogeneous magnetization there is
only the stray-field energy. For in-plane states, magnetic poles arise
exclusively at the side-face $S$ (i.e.\ $r\equiv R$), and we obtain for the
stray-field energy density
\beq\label{eqDipolarEnergy}
f^{\rm Dip} :=\frac{1}{V}\cdot\frac{J_s^2}{8\pi\mu_0}\,\int\limits_S
\int\limits_{S'}\,\frac{(\vec\al (\vec r)\cdot \vec n(\vec r)
)\cdot
(\vec\al (\vec r^{\,'})\cdot \vec n(\vec r^{\,'}))}{|\vec r
-\vec r^{\,'}|})\,{\rm d}^2S\,{\rm d}^2S^{\,'}\,\,.
\eeq
  By numerical evaluation, followed by a fit, we get:
 $f^{\rm Dip}\approx \frac{J_s^2}{8\pi\mu_0}\cdot
[2.41\cdot\ln (1+5.31\, (\frac{t_h}{2R})^{0.88}) ]$.

From these results one obtains for the critical thickness $t_c(R)$, where
the energies of the vortex state and the in-plane single-domain states
agree, with the ansatz (\ref{ansatz2}):
 \beq\label{eq10} t_c(R) \cong 2\,R\cdot\left \{
\frac{1}{5.31}\cdot\{\exp\left ( \frac{8\pi}{2.41} \cdot\frac{l_m^2}{R^2}\cdot [2.67
+\ln\frac{R}{5l_m}+0.5] \right )  - 1 \} \right\}^{\frac{1}{0.88}}  \,\,. \eeq
Here the term $\sim 0.5$ comes from the dipolar energy of the central part
of the vortex.

In Fig.~2, this result for the critical thickness is plotted against $R$
 (the solid line) and compared with the results of an unpublished simplified
 calculation, (see \cite{Hoffmann,Schneider2000}, the dashed line, i.e.\
 $t_{c,0}(R)$), where the fact that in the center of the vortex $\al_z$ is
 different from zero has been totally neglected (yielding a wrong
 asymptotics for $R\ll l_m$), and where the magnetostatic energy has been
 calculated by replacing the circular dot of radius $R$ and thickness $t_h$
 by a prolate ellipsoid. It turns out that this simple calculation is not
 bad and captures the essentials, although there is still a significant
 discrepancy w.r.t.\ the direct numerical simulation e.g.\ by the OOMMF
 program. In contrast the better results with our ansatz (\ref{ansatz2}) for
 the profile functions can hardly be distinguished, to the accuracy of the
 plot, from the results of the OOMMF simulation.

(Additionally, Metlov and Guslienko, in their recent paper \cite{Metlov} on
thicker cylindrical dots, have determined the transition line between
in-plane and out-of-plane single domain states, see Fig.\ 2 in
\cite{Metlov}. However for our flat dots with $t_h \ll R$ this addition is
not important. In fact, in our Fig.\ 2, the additional phase boundary would
be represented by a very steep line almost parallel to the abscissa axis,
which would terminate at our 'solid line' phase boundary at a thickness of
$\sim 43$ nm. This is far beyond the thicknesses we consider.)

\subsection{In-plane external fields}

For in-plane external fields, the vortex deforms, and its center shifts by an
amount $\Delta$ (in $x$-direction, if the external field is applied in the
$y$-direction; actually, see Fig.\ 8, below, the dynamic behaviour is more
complicated). For simplicity, in the following approximation we only take
into account the {\it shift}. Here both $\Delta < R$ and also $\Delta \ge R$
are considered, see below). With this approximation we minimize the
Landau-Lifshitz functional. For the 'stray-field energy density' we again
have to evaluate (\ref{eqDipolarEnergy}) with the (fictitious) magnetic
poles at the side-face $S$; additionally there is now the Zeeman energy
density

\beq f^{\rm Zeeman}=-J_s|H^{\rm ext}|\cdot\frac{\Gamma(R,\Delta ,l_m,c)}{\pi
R^2}\,, \eeq
\vglue 0.2 truecm

 \noindent where $\Gamma(R,\Delta ,l_m,c)$ is a function of the parameters
$R$, $\Delta$, $l_m$ and $c$. Note that -- as already mentioned -- $\Delta$
can be larger than $R$, although one expects that the transition from the
vortex state to the in-plane single-domain state just happens in the region
$\Delta\approx R$.

In Fig.\ 3, again for a circular Py dot of $R=150$ nm and $t_h=4$ nm in an
external in-plane field of $H^{\rm ext}_y=100$ Oe, the energy densities
(dipole energy, Zeeman energy, exchange energy, and total energy) are
plotted as a function of the shift parameter $\Delta$. The equilibrium
position of the vortex center would correspond to the minimum of the total
energy, which appears at $\Delta\approx 0.4\,R$.  With increasing
$\Delta\,\,(\to R\,)$, in Fig.\ 3 there is an {\it energy barrier},
separating the in-plane single-domain state (i.e.\
$\Delta \gg R$) from the deformed vortex state (i.e.\ $\Delta < R$). However
this energy barrier depends on $H^{\rm ext}_y$, and with increasing $H^{\rm
ext}_y$ it decays to zero as shown in Fig.~4.

Fig.\ 4 yields quantitative results for the 'annihilation fields' of the
vortex state rsp.\ for the energy barrier separating the vortex state from
the in-plane single-domain state, if the in-plane external field is not
strong enough for the transition. Since the deformation of the vortex during
the shift is neglected in our approximation, this energy barrier is
overestimated. In any case, near 'rough boundaries' the actual behaviour
will be too complex for a simple calculation, see \cite{Cowburn2}.

In Fig.~5, we compare our calculations with experimental results for the
annihilation fields of flat circular Py samples, plotted over the double
reciprocal aspect ratio $2R/t_h$, see \cite{Schneider}. In our calculation
we have used the experimental values $A=1.3\times 10^{-6}$ erg/ccm,
$J_s=0.9676$ Tesla, zero anisotropy, and two thicknesses ($t_h=15$ nm and
$t_h=8.3$ nm). As can be seen from the figure, in agreement with the
above-mentioned arguments, our theoretical estimates are systematically
$\approx 20-30$ {\%} too high, but a large part of the deviation is probably
due to the fact that in the experiments a non-neglegible perpendicular field
component was present, \cite{Rahm}; moreover, since the experiments are at
room temperature, the influence of thermal fields may also have been
non-neglegible, see ref.\ \cite{Fraermann,Li}. In any case, in the paper of
Guslienko and Metlov, \cite{Guslienko2002}, who also perform a comparison
with the experiments of \cite{Schneider}, the theoretical results are
significantly overestimated at small values of $R/t_h$, contrasted with our
Fig.\ 5. The discrepancy is due to the fact that the authors of
\cite{Guslienko2002} use basically the simplified profile function of Usov
and Peschany, \cite{Usov}, which should not be applied for $R\lsim t_h$. But
globally, i.e.\ averaged over all values of $R/t_h$, in \cite{Guslienko2002}, the
agreement between the calculated vortex annihilation fields and the
 experiments  looks better than that one presented in our Fig.~5, in spite
of the drastic overestimation at small values.
\section{Dynamics} This topic is particularly emphasized in our paper; to use
both the {\it vorticity} and the {\it central polarisation} of  flat
circular dots for information processing one should be able to switch both
properties reliably and independently, and -- of course -- fast, i.e.\
significantly below ns time scales.

Unfortunately, the vorticity of the static magnetic state cannot be directly
 influenced experimentally with {\it circular} dots. However, M.\ Schneider
 {\it et al.} have shown how one can easily influence the vorticity of the
 magnetic state by using dots, which are not perfectly circular, cutting-off
 a segment of the structure, see \cite{Schneider2001}. This refers to static
 or quasi-static conditions.  {\it But in the following we consider fast dynamic
 changes, at first by time-dependent in-plane fields, and then by
 time-dependent perpendicular fields~}:

 \subsection{Time-dependent in-plane fields} \label{seInplaneDynamics}

  As usual, we consider a circular Py dot ($A=1.3\times 10^{-6}$ erg/ccm,
 $J_s=1.08$ T, neglegible anisotropy) with $R=150$ nm and $4$ nm thickness.
 Without external field, the dot is in a stable vortex state. At time $t=0$,
 an in-plane external field $H^{\rm ext}_y$ is applied,
 and the 'critical duration' $t_{\rm crit}$ is determined, i.e.\ the time
 until the vortex has been shifted to the edge of the dot ($\Delta =R$), so
that the vortex state is either 'lost' to an in-plane single-domain state,
$\Delta\ge R$, or restored ($\Delta\to 0)$ after the field $H_y^{\rm ext}$ is
switched off. The results (which have been obtained by computer simulations
 with the OOMMF code, \cite{OOMMF}) are presented in Fig.~6; they depend on
 Gilbert's damping parameter $\al_G$ in a rather subtle way (see below). At
 first, we see from this figure that to reach critical times below $t=1$ ns
 one needs quite strong external fields, namely $H^{\rm ext}_y \rsim 300$ Oe
 to get below 800 ps, and $\rsim 500$ Oe to get below 600 ps. Moreover, the
 information contained in the polarisation of the vortex center is only
 preserved for unrealistically large $\al_G$ (e.g.\ for $\al_G=1$ the vortex
 center is only stable for fields $H_y^{\rm ext} < 1100$ Oe, i.e.\ up to the
 end-point of the lowest curve in Fig.~6.) For $\al_G=0.5$ the corresponding
  (second-lowest) curve in Fig.~6 ends already at $H_y^{\rm ext}=650$ Oe
with $t_{\rm crit}=500$ ps, and for $\al_G=0.01$ we have only obtained one
point in Fig.~6, namely with $t_{\rm crit}=1000$ ps and
 $H_y^{\rm ext}=200$ Oe. If the vortex state is 'lost' to an in-plane
 single-domain state by the application of the in-plane field $H_y^{\rm
 ext}$, one would need roughly at least another 'critical duration' to
 generate a new vortex state beginning from the boundary of the circular dot
 and propagating the core of
the new vortex to the center; maybe this new vortex has inverted vorticity,
inverted according to the systematic but slow prescription of
 \cite{Schneider2001,Usov2002}. In any case, this vorticity-switching would
be too slow, and too complex.

 In fact, one sees from Fig.~7, where in an ($\al_G$, $H^{\rm
 ext}_y$)--plane the boundary line separating stable and unstable regions of
 the (deformed) vortex state is plotted, that for realistically small values
 of $\al_G$ the vortex center is unstable for fields above $200$ Oe, i.e.\
 the independent information contained in the {\it central polarisation} of
 the vortex gets lost by the application of $H_y^{\rm ext}$: Stability of
 this 'polarisation information' can only be obtained for smaller in-plane
 fields $H_y^{\rm ext}$, which also means 'critical times' larger than 1 ns.

 In the next figure, Fig.~8, which corresponds to some kind of '{\it movie
sequence}', the particular case of $\al_G=0.02$ is considered in detail, for
$H_y^{\rm ext}=300$ Oe. (The observation that the vortex core moves in the
'north-east' direction, although from the fact that the switching-field is
applied in the y-direction one would expect that it should move exactly
'eastwards', comes from the following: In the first $\sim 100$ picoseconds
the vortex core, in fact, starts moving to the right, but at the same time
some kind of $180^{o}$ domain wall is formed on the segment between $x=0$
and $x=R$; under the influence of the $H_y$-field this domain wall tends to
move upwards, and after 600 ps it can still be identified in Fig.\ 8 to the
r.h.s. of the vortex core.  In any case, since the 'domain wall segment' is
fixed to the (slower) vortex core, the ''north-east'' motion of the core
results, as a consequence of the complicated coupling between all spins in
course of the dynamics. Further details are only seen if the 'movie'
corresponding to Fig.\ 8 is observed on a shorter time-scale.)

The total simulation time of  Fig.\ 8 is $1.5$ ns: One concludes that
vorticity annihilation is interesting and complicated, but rather slow
($t_{\rm crit}\approx 1.5$ ns); particularly, the information contained in
the central polarisation is lost by applying the field
$H_y^{\rm ext}$ for the critical duration, and one is far from simply
'switching' the vorticity. (These conclusions would also apply for larger
values of $\al_G$, although for $\al_G=1$, which is unrealistically large,
one would gain a factor of 3 in speed.)

As a consequence, for vorticity-switching one should not  use in-plane
fields, but instead one should work with perpendicular fields, as discussed
in the following.

\subsection{Time-dependent perpendicular fields}

 We have finally performed  simulations, starting from the planar
equilibrium vortex configuration, with pulsed external 'switching fields'
 which are now {\it perpendicular} to the plane. We consider the same
structure as before. The applied field pulse was very strong, $B_z^{\rm
ext}=\mu_0H_z^{\rm ext} =0.5$ T, i.e.\ $H_z^{\rm ext}\hat =5000$ Oe, but
also very short (see below). The Gilbert damping was assumed to be as small
as in Fig.\ 8, i.e.\ $\al_G=0.02$, but the pulse time of $H_z^{\rm ext}$ was
only $42$ ps, i.e.\ only a small fraction of the total simulation time of
the corresponding 'movie sequence', which we calculated for that example.
This time the vorticity is really inverted, and the switching is very fast,
42 ps, and without ''ringing'' after the end of the field-pulse.

 In Fig.~9 we consider the effective 'switching time' $t_{\rm
sw}=t_{\frac{1}{2}}$, i.e.\ the time needed to turn around the spins -- in a
spatial average -- by $180$ degrees, systematically as a function of
$H_z^{\rm ext}$. For vanishing damping this corresponds to a gyromagnetic
precession by an angle $\pi$, i.e.\
$t_{\frac{1}{2}}=\frac{\pi}{|\gamma_0|\cdot H^{\rm ext}_z}$. This is
shown in Fig.\ 10, which presents results from the first part of our
'movie'. After a duration of $t_{\frac{1}{2}}$, 42 ps in our case, the
field is switched off; the system then 'wiggles' around the new equilibrium
vortex state and gets slowly into equilibrium  after several ns. {\it But the
'wiggles' are small enough that the new equilibrium state can be clearly
recognized already after $42$ ps, although the system is still far from
equilibrium}, \cite{REMwiggling}. 

(Note that in Fig.\ 10 the exchange energy does not change with time, in
contrast to the magnetostatic energy, although after $\sim 21$ ps the
transient spin configuration has changed from a vortex state to a 'hedgehog
state'; this invariance of the exchange energy is in agreement with a
'topological statement' of W.\ D\"oring, \cite{Doering}.)

In Fig.~11, the 'wiggling' of the spatial average of $\al_z$ is presented.
Finally in Fig.~12 we plot, what happens to the polarisation of the vortex
center under the influence of a 'tilted vertical field pulse', involving a
strong perpendicular component $H_z^{\rm ext}=-1000$ Oe, applied between
$t=0$ and $60$ ps, accompanied by a simultaneous weaker in-plane pulse of
$H_y^{\rm ext}=150$ Oe between $t=0$ and $t=30$ ps. We find that at first
the central polarisation is inverted from (+) to (-), but
then it returns to the starting state (+).  In fact, we find it difficult to
control the central polarisation independently and with similar speed as the
vorticity.

That very strong  vertical fields in the kOe range are necessary for
polarity-switching of the vortex core is also found in a recent letter of
Kikuchi {\it et al.}, \cite{Kikuchi}.

 {\it We have seen that the fast switching of the {\it vorticity} by
vertical fields works successful, but the 'switching field pulses' are in
fact very large in strength, $\approx 5000$ Oe, and very short in duration,
$\approx 40$ ps. However, concerning the independent switching of the {\it
polarisation} of the center of the vortex, we always have found (for
realistically small values of $\al_G$) that the {\it central polarisation}
-- in contrast to the {\it vorticity} -- can hardly be controlled even by
strong field pulses oriented perpendicular to the plane; therefore in this
case we are unsuccessful and do not plot the results of our simulations.}
\newpage
\subsection{Vortex structure in flat circular dots: discussion}
Here we collect our results for the flat circular Permalloy (Py) dots considered
in this section 4:
 \ben

 \item The switching fields are quite large, i.e.\ necessary values are
higher than 300 Oersteds (for in-plane fields), and 5000 Oersteds for the
case of perpendicular fields.

\item For realistic values of the Gilbert damping parameter $\al_G$, see
\cite {Gilbert}, only the {\it vorticity} can be reliably
switched, \cite{REMgilbert}, but not the {\it central polarisation} of the
vortex core.

\item The time scale necessary for the precessional switching of the
vorticity is very short, $\tau\sim 40$ ps, for the case of perpendicular
field pulses. In that case, after the switching the magnetization ''wiggles''
slightly, i.e.\ there are relatively small oscillations, below
$\pm 10$\%, around the new
equilibrium.  These small precessional oscillations  decay rather slowly on
a much larger ns time scale.  {\it But principally the information contained
in the new state can already be obtained after 40 ps with sufficiently
strong perpendicular field pulses without necessity to suppress the
'wiggling'}, \cite{REMringing}.

\item For the case of in-plane switching fields the  field strengths for the
dynamic annihilation of the vortex state are an order of magnitude lower,
$B^{\rm ext}_y\sim 300$ Oersteds, and they need not to be pulsed; but the
time scales for vortex-switching with in-plane fields are also an order of
magnitude slower, namely $\rsim 500$ ps, \cite{Hoellinger}.

\een

 \section{Vortex structure in flat circular 'nano-rings'}
 As a consequence, at the end, we dismiss the vortex center totally,
although this implies the definite loss of the the possibility, to
manipulate a second bit. We consider flat circular 'nano-rings' with an
inner radius $R_1$ and an outer radius $R_2$. In this case, see Fig.~13,
with increasing $R_1$ the energy density of the vortex state decreases,
while that of the homogeneous in-plane state ('single-domain state')
increases significantly, i.e.\ the stability of the vortex state is
enhanced.

At the same time the switching of the vorticity by perpendicular field
pulses is hardly influenced by the size of $R_1$: In Fig.~14 we present
simulations for Py dots with $R_2=150$ nm and $t_h=4$ nm, comparing our
former results for $R_1=0$ with similar 'ring results' for $R_1=50$ nm, as a
function of the switching field $H_z^{\rm ext}$. Obviously the switching
time $t_{\frac{1}{2}}$ is practically not changed.

 \section{Conclusions}

\noindent Within the framework of the Landau-Lifshitz-Gilbert equation,
using permalloy parameters, we have studied the statics and dynamics of flat
circular magnetic nano-structures with an in-plane magnetic vortex
configuration putting particular emphasis on the (perpendicular) {\it
polarisation} of the vortex center (which may be shifted with respect to the
center of the circle), and on the (in-plane) {\it vorticity} of the magnetic
state. Studying switching processes induced by in-plane and out-of-plane
field pulses, we find that it is possible to switch the {\it vorticity} of
the magnetic dot on an ultra-short time scale of $40$ ps with strong enough
and short enough perpendicular external field pulses (strength
$\mu_0H_z^{{\rm ext}}\approx 0.5$ T; duration $\approx 40$ ps); but for
realistic values of the Gilbert damping constant $\al_G $, only the {\it
vorticity} can be switched this fast, but not the magnetic polarisation
$\al_z=\pm 1$ of the vortex core, so that it is better to dismiss the center
of the circular structures totally, concentrating instead on flat
nano-rings, i.e.\ with an inner radius $R_1$ and an outer radius $R_2$. In
such nano-rings, the stability of the vortex state is enhanced, and
concerning the switching of the vorticity, they have similar properties as
circular ones, i.e.\ with $R_1=0$.

 For in-plane fields, the field strength need only to be $\rsim 200$
Oer\-steds, but the time scales of vorticity-switching are $\rsim 1$ ns; the
vortex center is typically unstable, i.e.\ the topology of the state is
changed from a 'deformed vortex state' to a 'deformed in-plane single-domain
state', if in-plane fields of the 'critical duration' are applied. So the
case of 'in-plane vortex switching fields' should be dismissed rightaway for
circular dots.

 \section*{Acknowledgements} We acknowledge stimulating discussions with
 C.\ Back, G.\ Bayreuther, H.\ Hoffmann, M.\ Schneider, and J.\ Zweck.

\newpage

\centerline{\underline{\bf Figures and Figure Captions}}
\vglue 0.5 truecm

\epsfxsize=12cm
\epsfbox{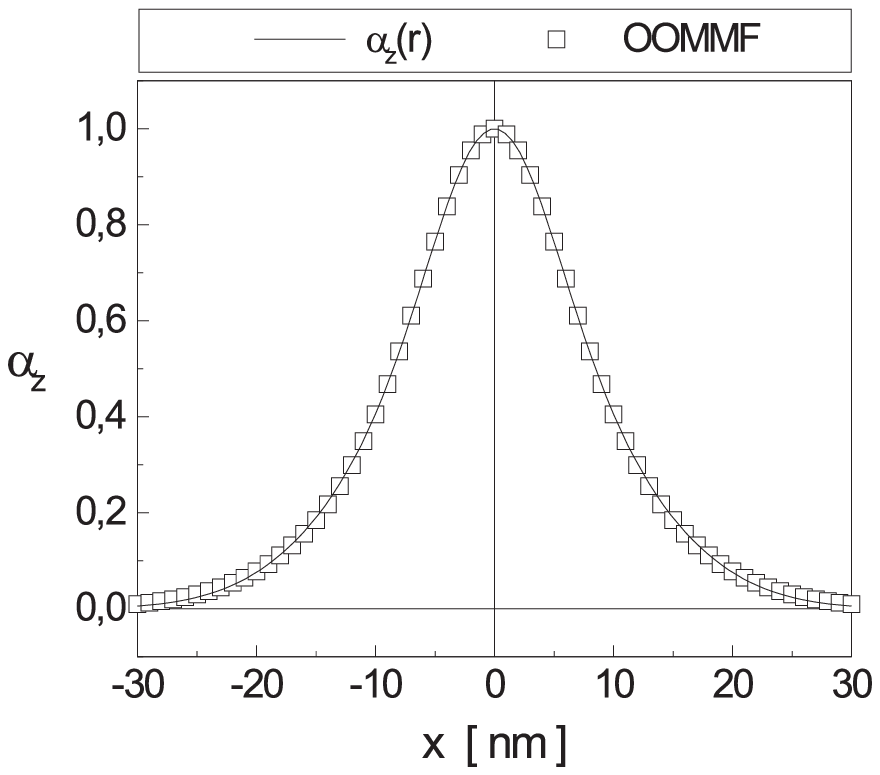}

\noindent{\bf Fig.1a:}
{{{The 'profile function' $\al_z(r)$ as given by our analytical ansatz is
compared with results of the direct numerical simulation by the OOMMF code.
Material parameters: Permalloy. The radius of the circular dot
considered is 175.5 nm, the height is 6 nm.}}}

\newpage

\epsfxsize 12cm
\epsfbox{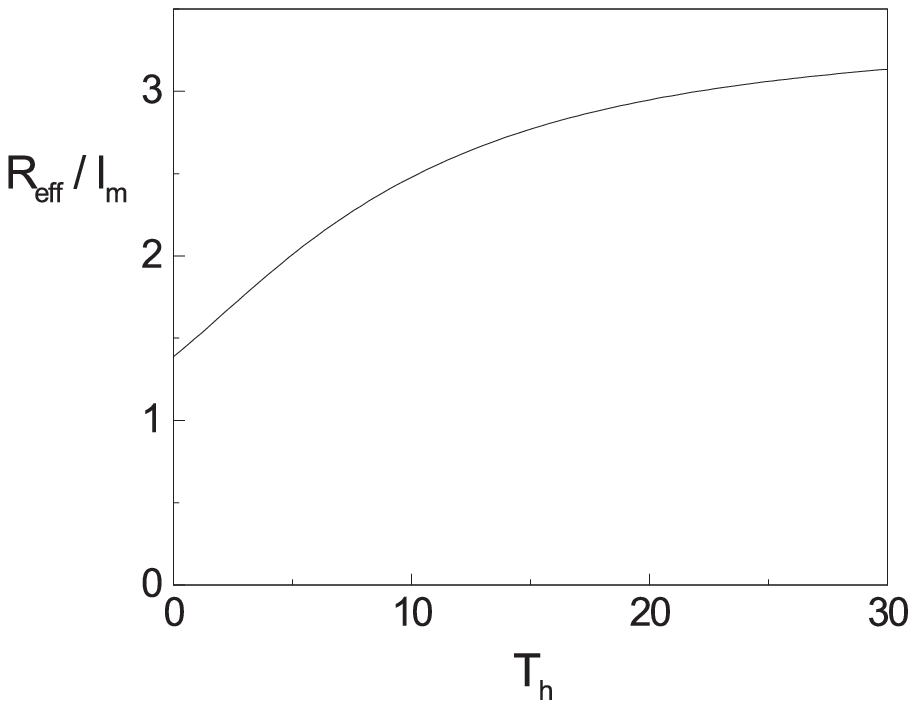}
\vglue 0.5 truecm

\noindent{\bf Fig.1b:}
{{{The effective radius $R_{\rm eff}$ of the vortex core is presented over
the reduced thickness $T_h :=\frac{t_h}{l_m}$ of the circular dot, in units
of the magnetic exchange length  $l_m$. Material parameters: Py, as in Fig.\ 1a.}}}

\newpage


\epsfxsize=12cm
\epsfbox{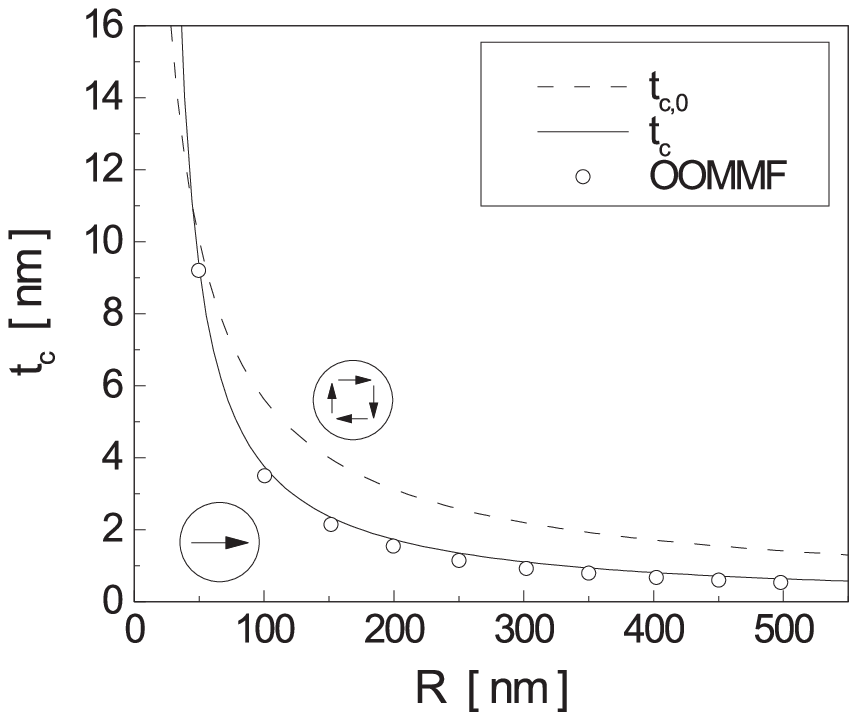}

\noindent {\bf Figure 2:}
{{{The critical thickness $t_c(R)$, i.e.\ the line which separates the
stability ranges of the vortex state from that of the in-plane single-domain
state, is presented as a function of the radius $R$ of the circular dot.
Parameters of Permalloy material are used. The solid line describes our
results, as given in the text; the dashed line presents the simplified
approximation
$t_{c,0}(R)$, also described in the text, and the circles describes the results
of a numerical calculation with the OOMMF code. (Besides, the additional
phase transition line separating the out-of-plane single-domain state from
the in-plane single domain state would in the present plot almost coincide
with the abscissa axis.)}}}
\newpage


\epsfxsize=12cm
\epsfbox{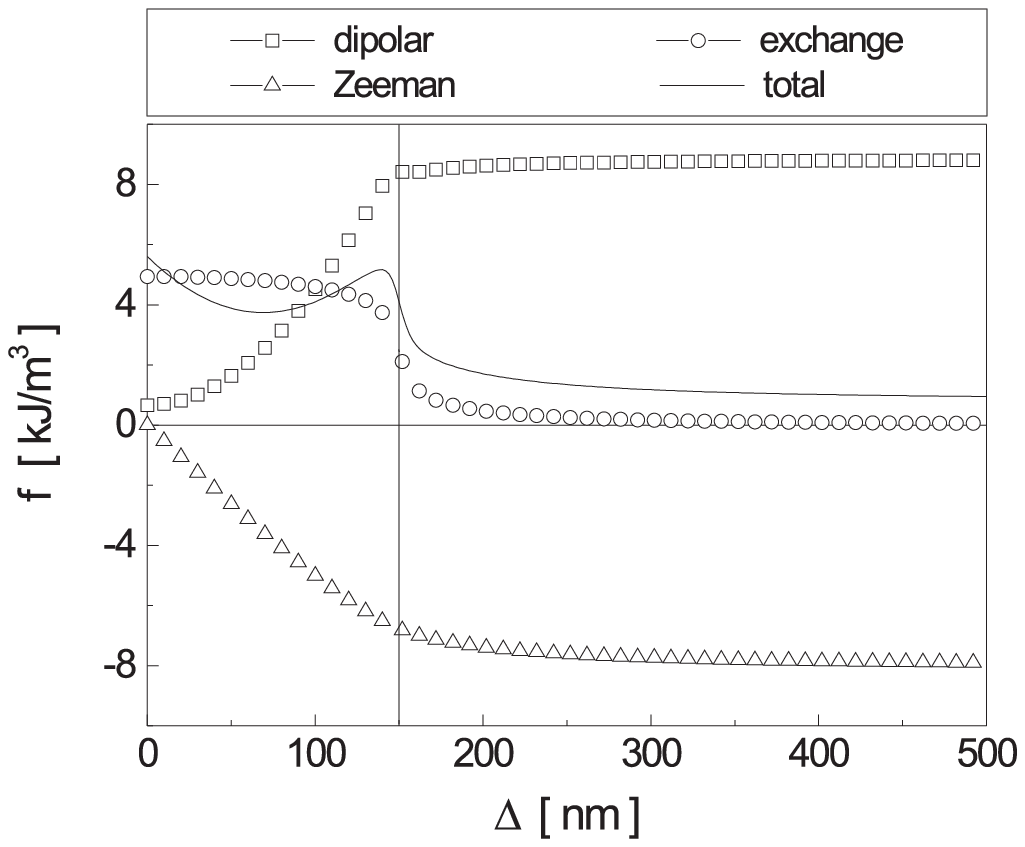}

\noindent {\bf Figure 3:}
{{{The total energy density $f$ ( the solid line), and the partial energy
densities of the dipolar energy (squares), exchange energy (diamonds) and
Zeeman energy (triangles)) for the vortex state in a circular permalloy dot
of radius $R=150$ nm (note the vertical line) and thickness $t_h=4$ nm are
presented as a function of the shift $\Delta$ of the vortex core;
 for more details see the text}}}

\newpage


\epsfxsize=12cm
\epsfbox{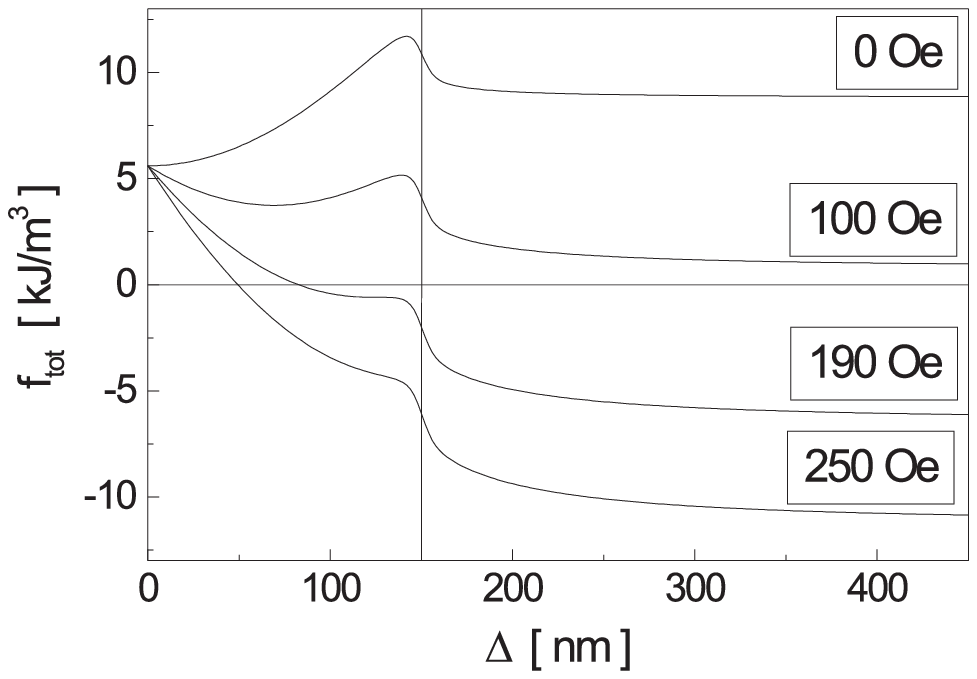}

\noindent {\bf Figure 4:}
{{{For different values of the external in-plane Zeeman field the density
$f_{\rm tot}$ of the total energy
of Fig.\ 3 is presented against the shift $\Delta$ of the vortex core.
One can see that for strong enough Zeeman fields ($H_y^{\rm ext} \ge 190$
Oe) there is no longer a local
minimum, i.e.\ the vortex vanishes ($\Delta\to\infty$), which means a transition
to a single-domain state. The vertical line denotes the radius $R$ of the
circular dot.}}}


\epsfxsize=12cm
\epsfbox{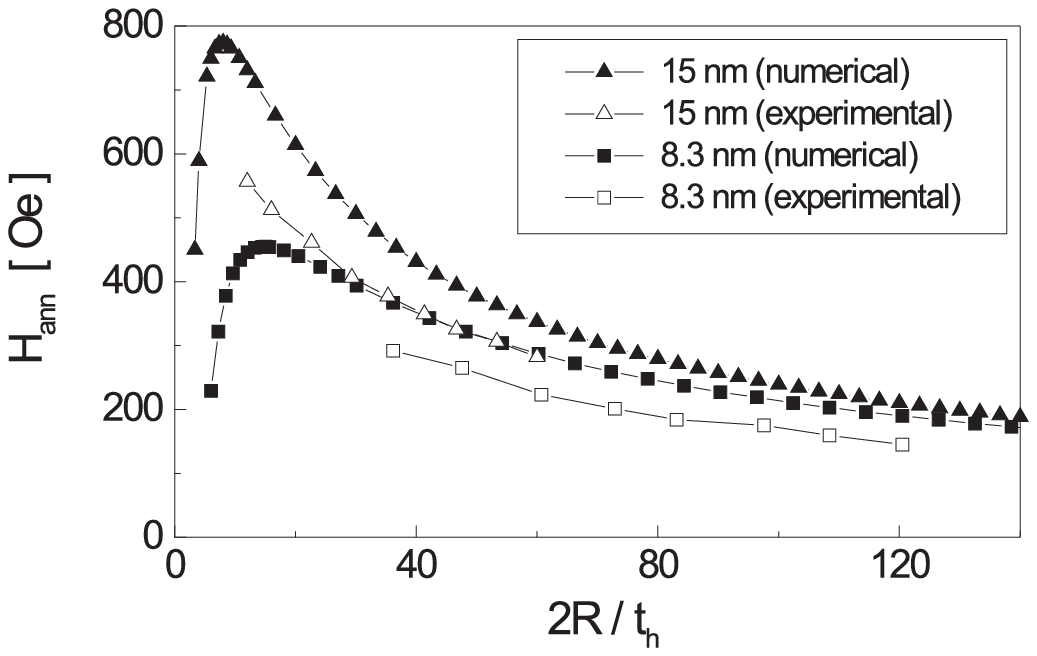}

\noindent {\bf Figure 5:}
{{{For Py dots of thickness $t_h=15$ nm and $8.3$ nm, respectively,
experimental results, and our theoretical results, for the in-plane annihilation
fields of the vortex state are presented over the double reciprocal aspect
ratio $2R/t_h$ of the dot.}}}

\newpage


\epsfxsize=12cm
\epsfbox{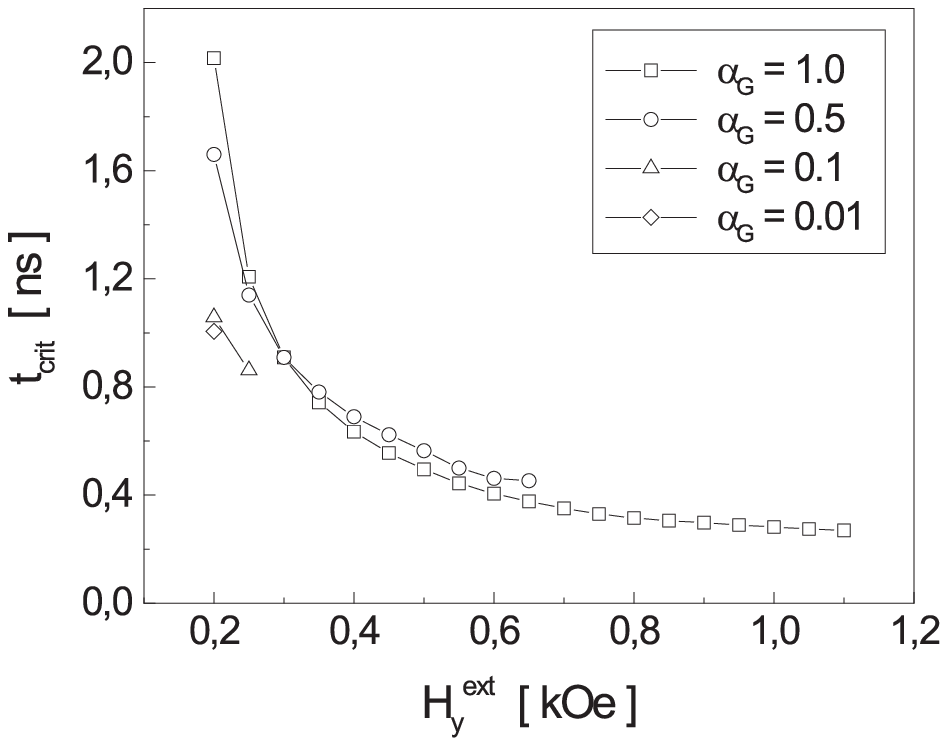}

\noindent {\bf Figure 6:}
{{{The critical time $t_c$ for dynamic annihilation of the vortex state
by in-plane fields $H_y^{\rm ext}$ is presented against $H_y^{\rm
ext}$ for different values of the Gilbert damping $\al_G$. For further
details see the text.}}}

\newpage


\epsfxsize=12cm
\epsfbox{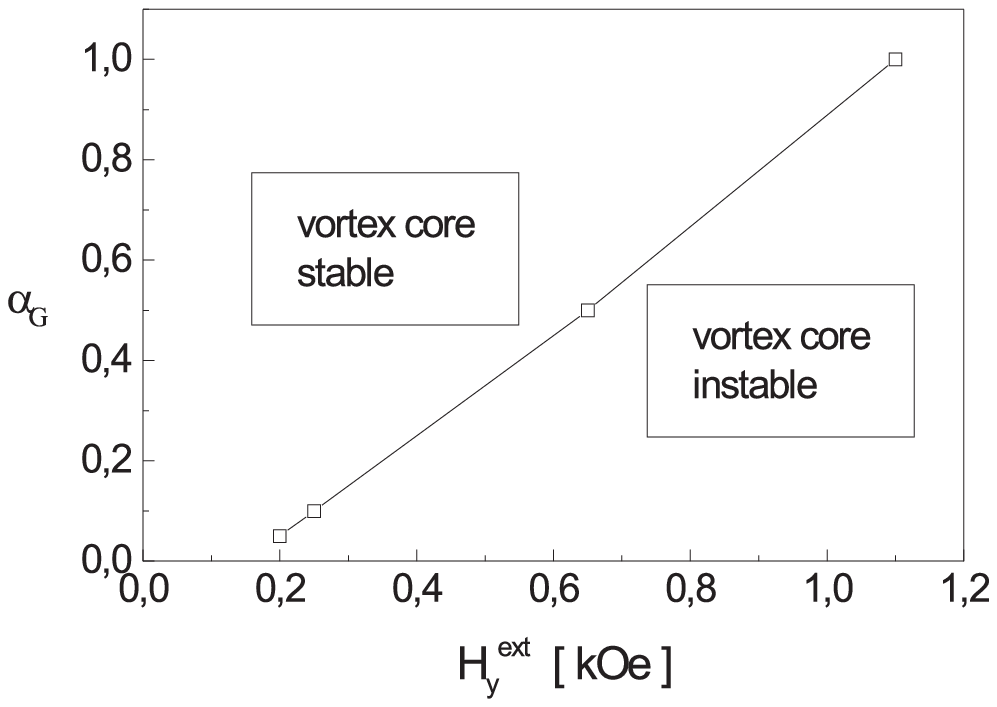}

\noindent {\bf Figure 7:}
{{{ The line separating the dynamic stability and instability regions
for dynamic annihilation of the vortex core by in-plane fields, as in the
preceding figure, is presented in an $(\al_G,H_y^{\rm ext})$ plane, where $\al_G$ is
Gilbert's damping parameter and $H_y^{\rm ext}$ the in-plane 'switching field'. 
}}}

\newpage


\epsfxsize=12cm
\epsfbox{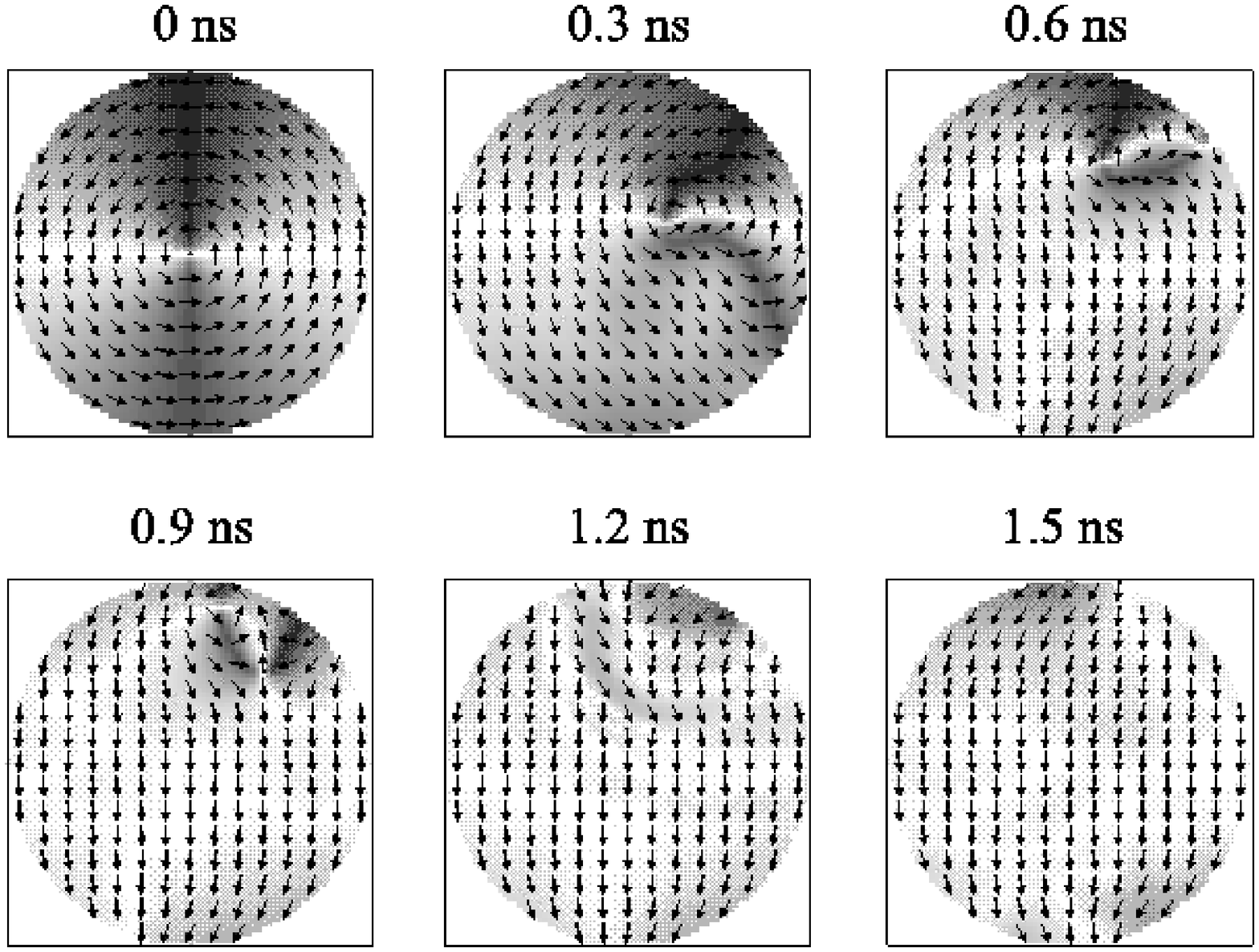}

\vglue 0.5 truecm
\noindent {\bf Figure 8:}
{{{ This 'movie sequence' presents the annihilation of a vortex state by
an in-plane field. This is a rather slow process, which needs at
least 1.5 ns, although the in-plane field is as strong as 300 Oe .  Gilbert's
damping parameter has been assumed to be $\al_G= 0.02$}}}

\newpage


\epsfxsize=12cm
\epsfbox{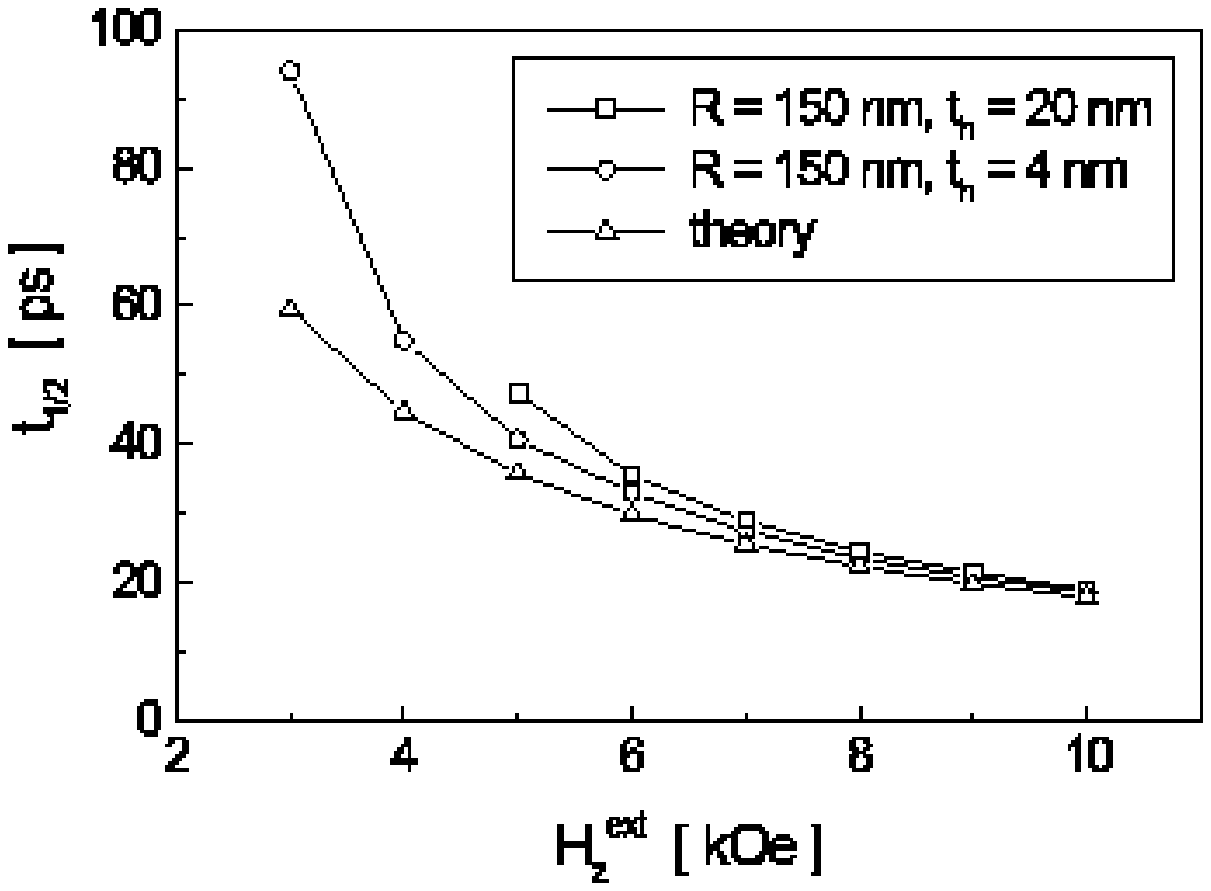}

\vglue 0.5 truecm
\noindent {\bf Figure 9:}
{{{ This plot presents the typical 'switching time' $t_{1/2}$ of a vortex
state by a pulsed perpendicular field $H_z^{\rm ext}$. This is a very fast
process, which needs typically only $\sim$ 40 ps, but the necessary strength
of the field is very large, $\sim$ 5000 Oe.  Gilbert's damping parameter has
been assumed to be $\al_G= 0.02$. ''Theory'' means the simple result
$t_{1/2} \,=\,\pi/(|\gamma_0| H_z^{\rm ext})$; the other results are from
OOMMF-simulations.}

\newpage


\epsfxsize=12cm
\epsfbox{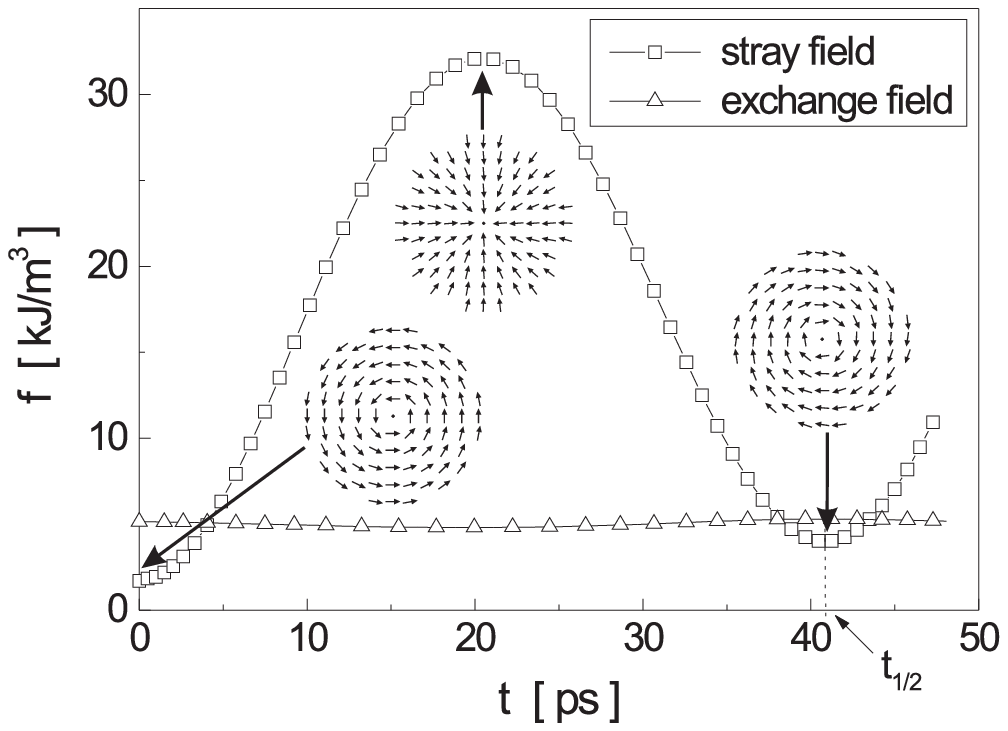}

\vglue 0.3truecm
\noindent {\bf Figure 10:}
{{{ This plot shows  details of the fast switching of the vorticity of our
flat circular Permalloy dot of radius $R=150$ nm and thickness $t_h=4$ nm by
a pulsed vertical field of $H_z^{\rm ext}=5000$ Oe. The energy densities $f$
of the dipolar energy ('stray-field') and of the exchange energy are
presented over the time for the first 50 ps after the start of the
perpendicular field pulse, which lasts from $t=0$ to $t=42$ ps.
}}}

\newpage


\epsfxsize=12cm
\epsfbox{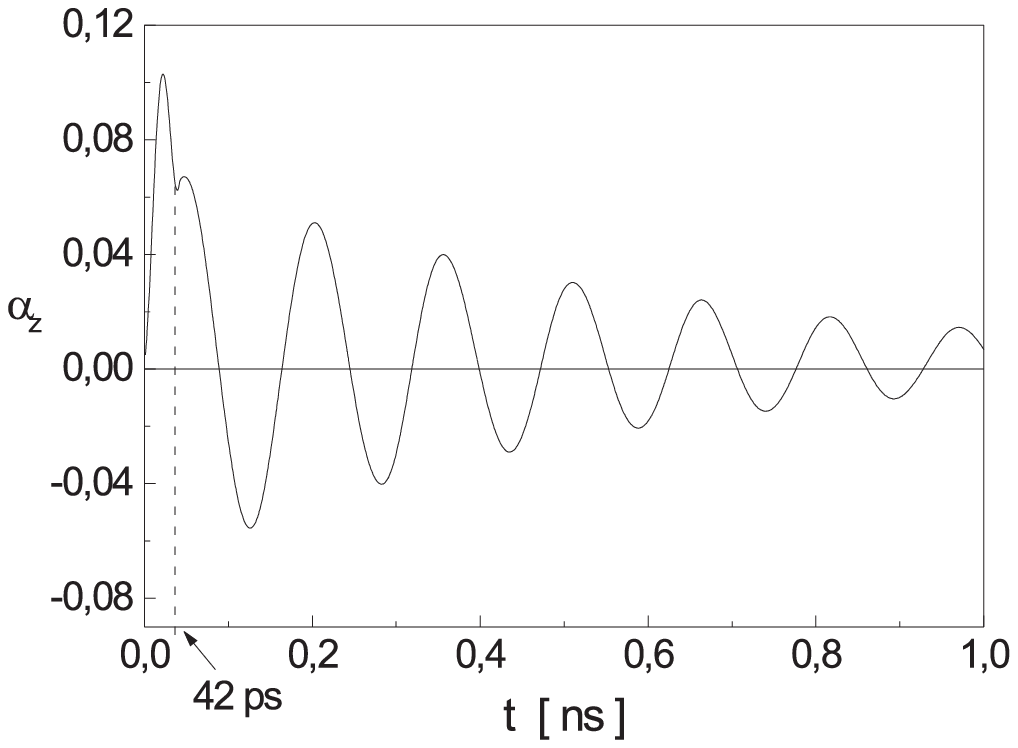}

\vglue 0.5 truecm
\noindent {\bf Figure 11:}
{{{ This plot shows the 'wiggling' of the small out-of-plane component
$\al_z$ of the (spatially averaged) magnetization of the circular Permalloy
dot; $\al_z$ comes to rest only after several ns, although the switching of
the vorticity needs essentially only 42 ps, see the preceding figure. Again
we have assumed
$\al_G=0.02$.
}}}

\newpage


\epsfxsize=12cm
\epsfbox{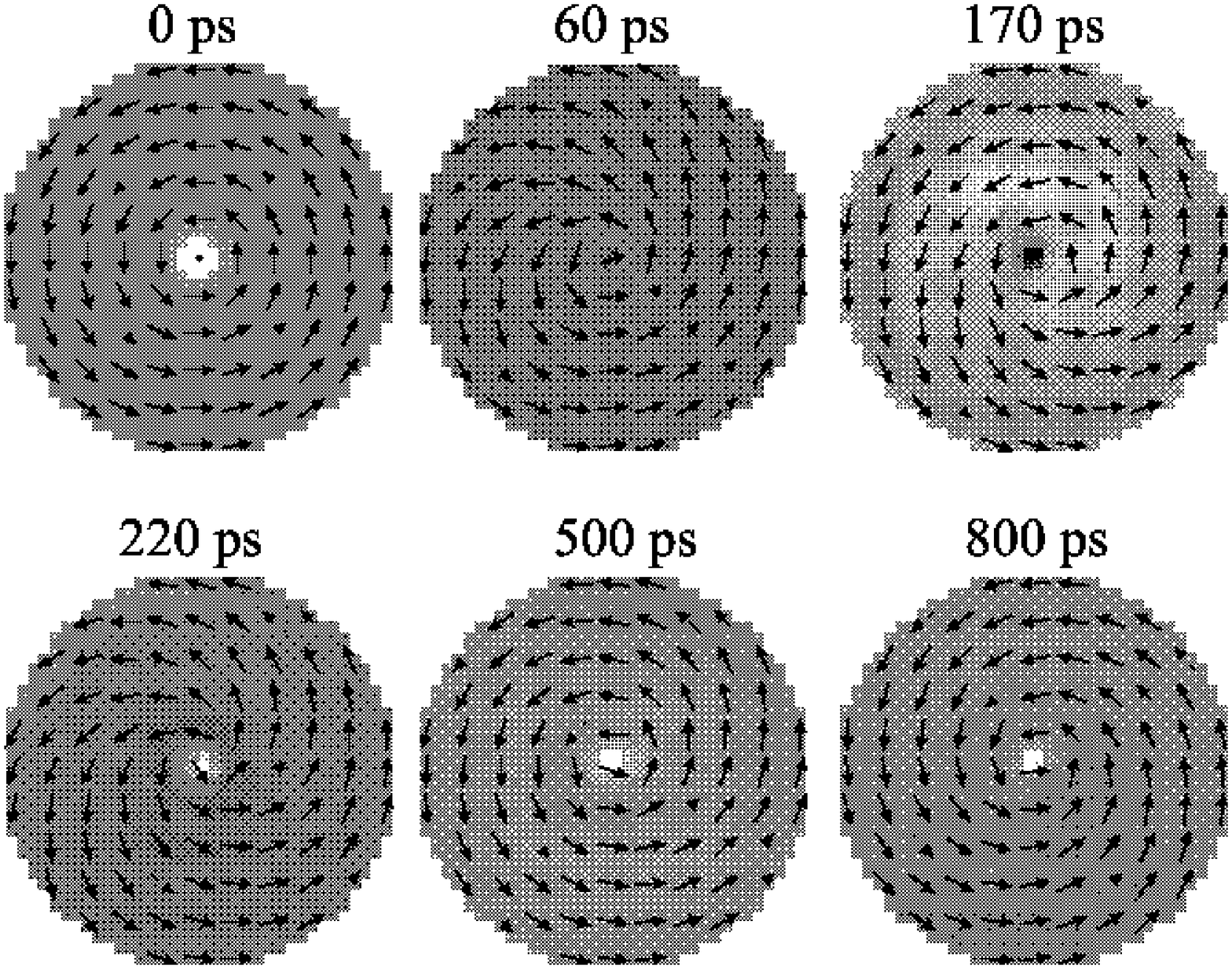}

\vglue 0.9 truecm

\noindent {\bf Figure 12:}
{{{Here white (black) background colours mean positive (negative) values of
$\al_z$; so this 'movie sequence' shows that under the conditions described
in the text the vorticity is not switched at all, whereas the central
polarity changes from +1 to -1 at $t\approx 60$ ps, remains there until
$t\approx 170$ ps,
but returns to +1 for $t \rsim 220$ ps. The applied field pulses were
$H_z^{\rm ext}=1$ kOe between 0 and 60 ps, and simultaneusly $H_y^{\rm ext}=
150$ Oe between 0 and 30 ps. As before, the radius and the thickness of our
Py dot were $150$ and $4$ nm, respectively, and the Gilbert damping was
$\al_G=0.02$. Note that for 60 ps and 220 ps one is no longer dealing with
a vortex state, but rather with a spiral.}}}

\newpage


\epsfxsize=12cm
\epsfbox{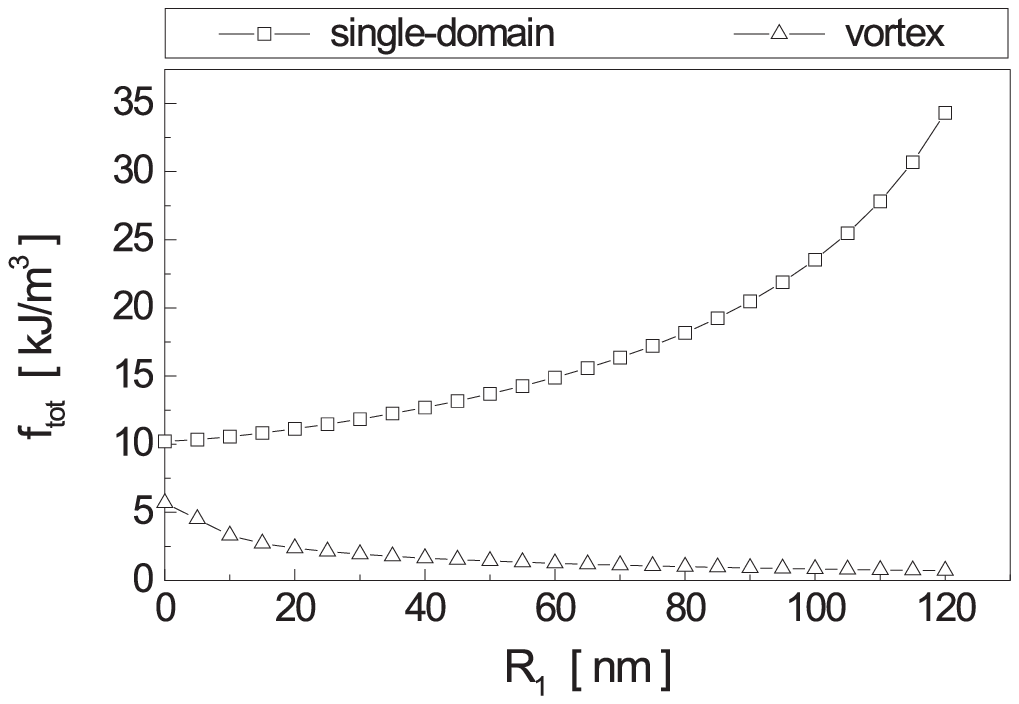}

\vglue 0.5 truecm
\noindent {\bf Figure 13:}
{{{For flat 'nano-rings' made of Py material with fixed outer radius
$R_2=150$ nm and thickness $t_h=4$ nm, the total energy densities $f_{\rm
tot}$ of the vortex state and the homogeneous in-plane state ('single-domain
state') are presented against the inner radius $R_1$.
}}}

\newpage


\epsfxsize=12cm

\epsfbox{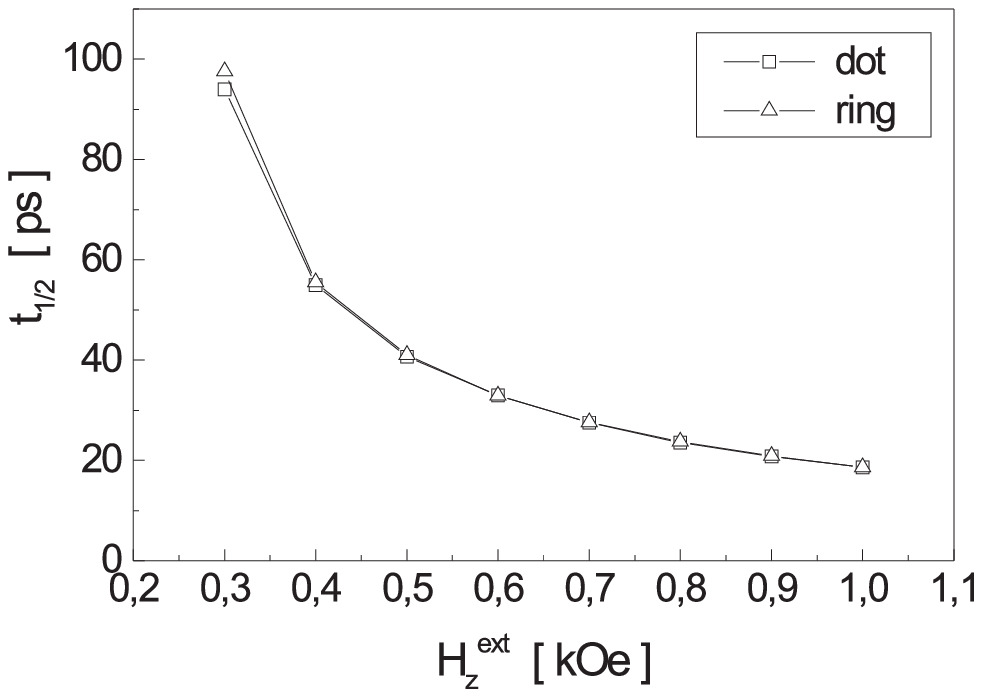}

\vglue 0.5 truecm
\noindent {\bf Figure 14:}
{{{ The switching time $t_{1/2}$ of the vorticity of a flat circular
Permalloy dot (thickness $t_h=4$ nm, radius $R=150$ nm) and of a
corresponding 'nano-ring' with outer radius $R_2=R$ and inner radius
$R_1=R/3$ is plotted against the vertical switching field $H_z^{\rm ext}$.
}}}
\newpage

\centerline{\underline{\bf Figure Captions}}
\vglue 0.5 truecm

\noindent{\bf Fig.1a:}
{{{The 'profile function' $\al_z(r)$ as given by our analytical ansatz is
compared with results of the direct numerical simulation by the OOMMF code.
Material parameters: Permalloy. The radius of the circular dot
considered is 175.5 nm, the height is 6 nm.}}}

\noindent{\bf Fig.1b:}
{{{The effective radius $R_{\rm eff}$ of the vortex core is presented over
the reduced thickness $T_h :=\frac{t_h}{l_m}$ of the circular dot, in units
of the magnetic exchange length  $l_m$. Material parameters: Py, as in Fig.\ 1a.}}}

\noindent {\bf Figure 2:}
{{{The critical thickness $t_c(R)$, i.e.\ the line which separates the
stability ranges of the vortex state from that of the in-plane single-domain
state, is presented as a function of the radius $R$ of the circular dot.
Parameters of Permalloy material are used. The solid line describes our
results, as given in the text; the dashed line presents the simplified
approximation
$t_{c,0}(R)$, also described in the text, and the circles describes the results
of a numerical calculation with the OOMMF code. (Besides, the additional
phase transition line separating the out-of-plane single-domain state from
the in-plane single domain state would in the present plot almost coincide
with the abscissa axis.)}}}

\noindent {\bf Figure 3:}
{{{The total energy density $f$ ( the solid line), and the partial energy
densities of the dipolar energy (squares), exchange energy (diamonds) and
Zeeman energy (triangles)) for the vortex state in a circular permalloy dot
of radius $R=150$ nm (note the vertical line) and thickness $t_h=4$ nm are
presented as a function of the shift $\Delta$ of the vortex core; for more
details see the text}}}

\noindent {\bf Figure 4:}
{{{For different values of the external in-plane Zeeman field the density
$f_{\rm tot}$ of the total energy
of Fig.\ 3 is presented against the shift $\Delta$ of the vortex core.
One can see that for strong enough Zeeman fields ($H_y^{\rm ext} \ge 190$
Oe) there is no longer a local
minimum, i.e.\ the vortex vanishes ($\Delta\to\infty$), which means a transition
to a single-domain state. The vertical line denotes the radius $R$ of the
circular dot.}}}

\noindent {\bf Figure 5:}
{{{For Py dots of thickness $t_h=15$ nm and $8.3$ nm, respectively,
experimental results, and our theoretical results, for the in-plane annihilation
fields of the vortex state are presented over the double reciprocal aspect
ratio $2R/t_h$ of the dot.}}}

\noindent {\bf Figure 6:}
{{{The critical time $t_c$ for dynamic annihilation of the vortex state
by in-plane fields $H_y^{\rm ext}$ is presented against $H_y^{\rm
ext}$ for different values of the Gilbert damping $\al_G$. For further
details see the text.}}}

\noindent {\bf Figure 7:}
{{{ The line separating the dynamic stability and instability regions
for dynamic annihilation of the vortex core by in-plane fields, as in the
preceding figure, is presented in an $(\al_G,H_y^{\rm ext})$ plane, where $\al_G$ is
Gilbert's damping parameter and $H_y^{\rm ext}$ the in-plane 'switching field'. 
}}}

\noindent {\bf Figure 8:}
{{{ This 'movie sequence' presents the annihilation of a vortex state by
an in-plane field. This is a rather slow process, which needs at
least 1.5 ns, although the in-plane field is as strong as 300 Oe .  Gilbert's
damping parameter has been assumed to be $\al_G= 0.02$}}}

\noindent {\bf Figure 9:}
{{{ This plot presents the typical 'switching time' $t_{1/2}$ of a vortex
state by a pulsed perpendicular field $H_z^{\rm ext}$. This is a very fast
process, which needs typically only $\sim$ 40 ps, but the necessary strength
of the field is very large, $\sim$ 5000 Oe.  Gilbert's damping parameter has
been assumed to be $\al_G= 0.02$.  ''Theory'' means the simple result
$t_{1/2} \,=\,\pi/(|\gamma_0| H_z^{\rm ext})$; the other results are from
OOMMF-simulations. }}}

\noindent {\bf Figure 10:}
{{{ This plot shows  details of the fast switching of the vorticity of our
flat circular Permalloy dot of radius $R=150$ nm and thickness $t_h=4$ nm by
a pulsed vertical field of $H_z^{\rm ext}=5000$ Oe. The energy densities $f$
of the dipolar energy ('stray-field') and of the exchange energy are
presented over the time for the first 50 ps after the start of the
perpendicular field pulse, which lasts from $t=0$ to $t=42$ ps.
}}}

\noindent {\bf Figure 11:}
{{{ This plot shows the 'wiggling' of the small out-of-plane component
$\al_z$ of the (spatially averaged) magnetization of the circular Permalloy
dot; $\al_z$ comes to rest only after several ns, although the switching of
the vorticity needs essentially only 42 ps, see the preceding figure. Again
we have assumed
$\al_G=0.02$.
}}}

\noindent {\bf Figure 12:}
{{{Here white (black) background colours mean positive (negative) values of
$\al_z$; so this 'movie sequence' shows that under the conditions described
in the text the vorticity is not switched at all, whereas the central
polarity changes from +1 to -1 at $t\approx 60$ ps, remains there until
$t\approx 170$ ps,
but returns to +1 for $t \rsim 220$ ps. The applied field pulses were
$H_z^{\rm ext}=1$ kOe between 0 and 60 ps, and simultaneusly $H_y^{\rm ext}=
150$ Oe between 0 and 30 ps. As before, the radius and the thickness of our
Py dot were $150$ and $4$ nm, respectively, and the Gilbert damping was
$\al_G=0.02$. Note that for 60 ps and 220 ps one is no longer dealing with
a vortex state, but rather with a spiral.}}}

\noindent {\bf Figure 13:}
{{{For flat 'nano-rings' made of Py material with fixed outer radius
$R_2=150$ nm and thickness $t_h=4$ nm, the total energy densities $f_{\rm
tot}$ of the vortex state and the homogeneous in-plane state ('single-domain
state') are presented against the inner radius $R_1$.
}}}

\noindent {\bf Figure 14:}
{{{ The switching time $t_{1/2}$ of the vorticity of a flat circular
Permalloy dot (thickness $t_h=4$ nm, radius $R=150$ nm) and of a
corresponding 'nano-ring' with outer radius $R_2=R$ and inner radius
$R_1=R/3$ is plotted against the vertical switching field $H_z^{\rm ext}$.
}}}


\begin{thebibliography}{99}
\bibitem{REM1} It is perhaps not generally known that in polarizable matter
the transient electrodynamics following a short perturbation  is determined
by retardation effects involving the vacuum velocity of light, and not by
the slower velocity of stationary light waves in that matter.
\bibitem{REMoutofplane} The out-of-plane single domain state, which would be
energetically very unfavourable, does not play a role for our flat permalloy
dots.
\bibitem{REMwiggling} This relative 'harmlessness' of the ''wiggling'' in the
vorticity switching  is in contrast to the 'severeness' of the
 ''ringing'' phenomenon in the precessional switching of single-domain
states, where ''ringing'' poses severe problems; see the contributions in
\cite{Gerrits,REMringing,Hoellinger}.
\bibitem{REMgilbert} The Gilbert parameter for Py is very small, $\al_G=0.008$, and at
the interfaces of thin films or trilayers this is only enhanced to typical
values around $a_G\approx 0.1$ or below, \cite{refTserkovnyak}.
\end{thebibliography}
\end{document}